\documentclass[usegraphicx,useAMS]{basi}

      \DeclareSymbolFont{UPM}{U}{eur}{m}{n}
      \DeclareMathSymbol{\umu}{0}{UPM}{"16}

\newcommand{\lapp}{\mbox{\raisebox{-0.3em}{$\stackrel{\textstyle <}{\sim}$}}}
\newcommand{\gapp}{\mbox{\raisebox{-0.3em}{$\stackrel{\textstyle >}{\sim}$}}}

\title[Associated  {\rm H}{\sc i} absorption towards the core of 3C\,321]
      {Associated H{\sc i} absorption towards the core of the radio galaxy 3C\,321}
\author[Yogesh Chandola et al.]
       {Yogesh Chandola$^1$$\thanks{E-mail: chandola@ncra.tifr.res.in }$, S.K. Sirothia$^1$,
                       D.J. Saikia$^{1,2}$ and Neeraj Gupta$^3$  \\
$^1$NCRA, TIFR, Pune University Campus, Post Bag 3, Pune 411 007, India \\
$^2$Cotton College State University, Panbazar, Guwahati 781 001, Assam, India\\
$^3$ASTRON, Oude Hoogeveensedijk 4, 7991 PD Dwingeloo, The Netherlands \\
        }

\date{Received 2012 May 23; accepted 2012 July 22}

\pagerange{\pageref{firstpage}--\pageref{lastpage}}
\pubyear{  }

\begin{document}

\maketitle

\label{firstpage}

\begin{abstract}
We report the results of Giant Metrewave Radio Telescope (GMRT) observations of 
H{\sc i} absorption towards the FRII radio galaxy 3C321 (J1531+2404), which is 
associated with an active galaxy interacting with a companion. 
The absorption profile towards the radio core is well resolved and consists 
three components, of which the two prominent ones are red-shifted by 186 and 235 km s$^{-1}$ relative 
to the optical systemic velocity. The neutral hydrogen column density towards the core is
estimated to be $N$(H{\sc i})=9.23$\times$10$^{21}$(${T}_{\rm s}$/100)($f_{c}$/1.0) cm$^{-2}$, where
${T}_{\rm s}$ and $f_c$ are the spin temperature and covering factor of the background source
respectively. We also present radio continuum observations of the source with both the
GMRT and the Very Large Array (VLA) in order to understand the properties of a plume
of emission at an angle of $\sim$30$^\circ$ to the source axis. This feature appears to
have a steep high-frequency spectrum.  The current hotspots and jet are active and seen
in X-ray emission. The spectral ages of the lobes are
$\lapp$26 Myr. We discuss the possibility that the  plume could 
be relic emission due to an earlier cycle of activity. 
\end{abstract}


\section{Introduction}
Studying the properties of the gaseous environments of radio galaxies and 
quasars  on different scales are important for understanding the interactions 
of the radio jets with the external environment and the evolution of these sources.
Such studies could also provide useful insights towards understanding the triggering of
radio activity, and examining consistency of these properties with the 
unified schemes for active galactic nuclei (AGN) (Pihlstr\"om, Vermeulen \& Conway 2003; 
Gupta \& Saikia 2006b).  A useful way of probing the cold
 neutral component of this gas is via 21-cm H{\sc i} absorption towards
radio sources, which range in size from the sub-galactic sized compact
steep-spectrum (CSS) and gigahertz peaked-spectrum (GPS) sources, to the
large radio galaxies and quasars which are up to a few Mpc in size. 
The CSS and GPS objects (O' Dea 1998) have been inferred to be young ($<$10$^5$ yr), while
the larger sources could be older than $\sim$10$^8$ yr (Jamrozy et al. 2008; Konar et al. 2008).
  
Several H{\sc i} absorption line studies have shown that CSS and GPS objects
tend to exhibit absorption lines more frequently than larger sources, with the lines being both
blue- and red-shifted relative to the systemic velocity and exhibiting complex
line profiles. The H{\sc i} column densities also appear to be anticorrelated 
with the source sizes (Pihlstr\"om, Conway \& Vermeulen 
2003; Gupta et al. 2006), and are broadly consistent with the unified scheme for
radio galaxies and quasars. The relationship between the H{\sc i} column density and core prominence, which is
being used as a statistical indicator of source orientation, appears to be consistent
with the H{\sc i} gas being distributed in a circumnuclear disk on a scale smaller 
than the size of the compact radio sources (Gupta \& Saikia 2006b). Evidence of 
absorption arising from a circumnuclear disk-like structure has also been inferred from 
higher-resolution, Very Long Baseline Interferometric (VLBI)-scale spectroscopic observations in several
sources such as the CSS object J0119+3210 (4C+31.04) and the cores of
a few larger sources, Cyg\,A, NGC\,4261 and Hydra\,A (e.g. Conway \& Blanco 1995; 
Taylor 1996; Conway 1999; van Langevelde et al. 2000).
A comparison of the H{\sc i} absorption properties towards the cores of larger sources
with those of the CSS and GPS objects might provide insights towards understanding the 
evolution of the gaseous properties as the source ages. Such a study indicates that
the detection of H{\sc i} in absorption towards the cores of larger sources is significantly
smaller than for CSS and GPS objects suggesting an evolution in the gaseous content
of the host galaxies with source age (Chandola et al., in preparation).    
There have also been suggestions that the torus/disk may be different in  
FR\,I and FR\,II sources, which is likely to be reflected in the H{\sc i} absorption
properties of these two classes of sources (e.g. Morganti et al. 2001). Given the small
number of detections at present, the difference in detection rate between  
FR\,I and FR\,II sources does not appear to be significant (Chandola et al., in preparation).

An important aspect in our understanding of AGN is the episodic nature of
their nuclear activity, and the physical processes that might be governing
it (Saikia \& Jamrozy 2009 for a review). There appears to be a trend for a high rate of
detection of H{\sc i} absorption in sources with evidence of rejuvenated activity.
These include the giant radio galaxy 3C236 which also
exhibits evidence of star formation (Conway \& Schilizzi 2000), the
giant radio galaxy J1247+6723 with a GPS core (Saikia, Gupta \& Konar  2007), the misaligned DDRG
3C293 (Beswick et al. 2004; Emonts et al. 2005), 
the well-studied southern radio galaxy Centaurus A (Sarma, Troland \& Rupen 2002; Morganti et al. 2008), 
4C~29.30 (Chandola, Saikia \& Gupta 2010) and CTA~21 (Salter et al. 2010).  
The well-known FRII radio galaxy
Cygnus A, which has been shown to have two cycles of radio activity
from radio and X-ray observations (Steenbrugge, Blundell \& Duffy 2008;
Steenbrugge, Heywood \& Blundell 2010), also exhibits 
nuclear H{\sc i} absorption (Conway 1999).  

In this paper we present H{\sc i} observations with the Giant Metrewave Radio 
Telescope (GMRT) to localise and put constraints on the size of the H{\sc i}
absorber in the FRII radio galaxy 3C321, which has been reported earlier from 
Arecibo observations (Mirabel 1990). We also present radio continuum observations 
with the GMRT 
and the Very Large Array (VLA) of the diffuse plume of emission to explore whether 
this feature might be relic emission from an earlier cycle of activity. 

\section{3C321}
The radio galaxy 3C321 is an FRII radio galaxy at a redshift of 0.0961 
(luminosity distance = 436 Mpc) so that 1 arcsec corresponds to 1.759 kpc
in a Universe with H$_0$=71 km s$^{-1}$ Mpc$^{-1}$, $\Omega_{\rm m}$=0.27 and 
$\Omega_{\rm vac}$=0.73. Our observations show that it has two prominent 
hot-spots at the outer edges with an overall angular size of $\sim$ 286 arcsec, which corresponds to a projected
linear size of $\sim$ 503 kpc, similar to that found in the literature 
(e.g. Baum et al. 1988; Lal, Hardcastle \& Kraft 2008). 
On smaller scales, a VLA image with an angular resolution of $\sim$1.4 arcsec by Baum et al. (1988),
shows a compact core, a knot towards the north-west separated from the core by  $\sim$3.5 arcsec (6 kpc), 
and a somewhat diffuse but bright and prominent jet extending upto $\sim$38 kpc. There is also evidence of a weaker 
counter-jet towards the south-east. A lower-resolution radio image has shown
evidence of collimated emission extending all the way to the north-western hot-spot,
and a diffuse plume of emission starting beyond the bright emission in the
jet and extending for $\sim$140 kpc at an angle of $\sim$30$^\circ$ to the source axis
(Evans et al. 2008). 

Optical observations show that the galaxy associated with the radio source
has a prominent kpc-scale dust lane, and a fainter companion towards the 
north-west (Hurt et al. 1999; Martel et al. 1999; Roche \& Eales 2000;
de Koff et al. 2000). The two galaxies are embedded in a common stellar
envelope, and the velocity difference between the two galaxies is within
$\sim$200 km s$^{-1}$. Optical spectroscopic and X-ray observations suggest
that the companion galaxy also hosts an AGN (Filippenko 1987; 
Robinson et al. 2000; Evans et al. 2008). There are different regions of 
X-ray emission, with emission from both the hot-spots and regions offset
from the radio peaks (Hardcastle et al. 2004; Evans et al. 2008). Hubble
Space Telescope (\emph{HST}) STIS NUV and [O{\sc {iii}}] images, along with 
the {\it Chandra} and {\it Spitzer} IRAC 4.5 ${\umu}$m images show the complex 
distribution of gas in the circumnuclear region (e.g. Evans et al. 2008).
 
%%%%%%%%%%%%%%%%%%%%%%%%%%%%%%%%%%%%%%%%%%%%%%%%%%%%%%%%%%%%%%
%%%%%%%%%%%%%%%%%%%%%%%%%%%%%%%%%%%%%%%%%%%%%%%%%%%%%%%%%%%%%%

\section{Observations and data reduction}
The results presented here are based on H{\sc i} observations
made with the GMRT, as well as on archival GMRT and VLA data.
The observing log for both the GMRT and VLA observations is 
listed in Table~1.

%%%%%%%%%%%%%%%%%%%%%%%%%%%%%%%%%%%%%%%%%%%%%%%%%%%%%%%%%%%%%%%%%%%%%%%%%%%%%%%%%%%%%%%%%%%%%%%
\begin{table}
\caption{The observation log. Column 1: the name of the telescope, and the array configuration for the VLA observations;
column 2: dates of the observations; columns 3 and 4: the frequency and bandwidth used in making the images.}
\begin{center}
\begin{tabular}{c c c c c c  }
\hline
Telescope & Observation & Observed   & Bandwidth \\ 
          &  date        & Freq.      &           \\
          &              &(MHz)       &(MHz)      \\
 (1)      &    (2)       & (3)        & (4)       \\
\hline    
GMRT$^a$&   2005 May 31 & 614    &  28   \\
GMRT    &   2009 Feb 01 & 1295   &  16    \\
VLA-C$^a$&   1986 Dec 02 & 1511   &  50     \\
VLA-D$^a$&   1990 Jan 08 & 4860   &  100    \\
\hline
\end{tabular}

$^a$ archival data   
\end{center}
\label{table1}

\end{table}
%%%%%%%%%%%%%%%%%%%%%%%%%%%%%%%%%%%%%%%%%%%%%%%%%%%%%%%%%%%%%%%%%%%%%%%%%%%%%%%%%%%%%%%%%%%%%%%%%

The GMRT H{\sc i} observations were made in the standard manner, with  each observation
of the target-source interspersed with observations of the phase calibrator (J1609+266).
3C286 and 3C48 were observed as the flux density and bandpass calibrators, and all flux
densities are on the Baars et al.(1977) scale using the latest VLA values. 
The source was observed in a full-synthesis run of approximately 8 hours including calibration overheads.
The GMRT H{\sc i} data, as well as the archival GMRT data were 
calibrated and reduced using the pipeline developed by one of us (SS). We also 
reduced the GMRT archival data at lower frequencies but found the image and data quality to be 
unsatisfactory and have not included these here. The archival VLA data were reduced in the 
standard way using the Astronomical Image Processing System (AIPS) package.

\section{Results and discussion}

\subsection{H{\sc i} absorption spectra}
\begin{figure}
\vbox{
\includegraphics[angle=0, totalheight=4in, viewport=19 212 573 627, clip]{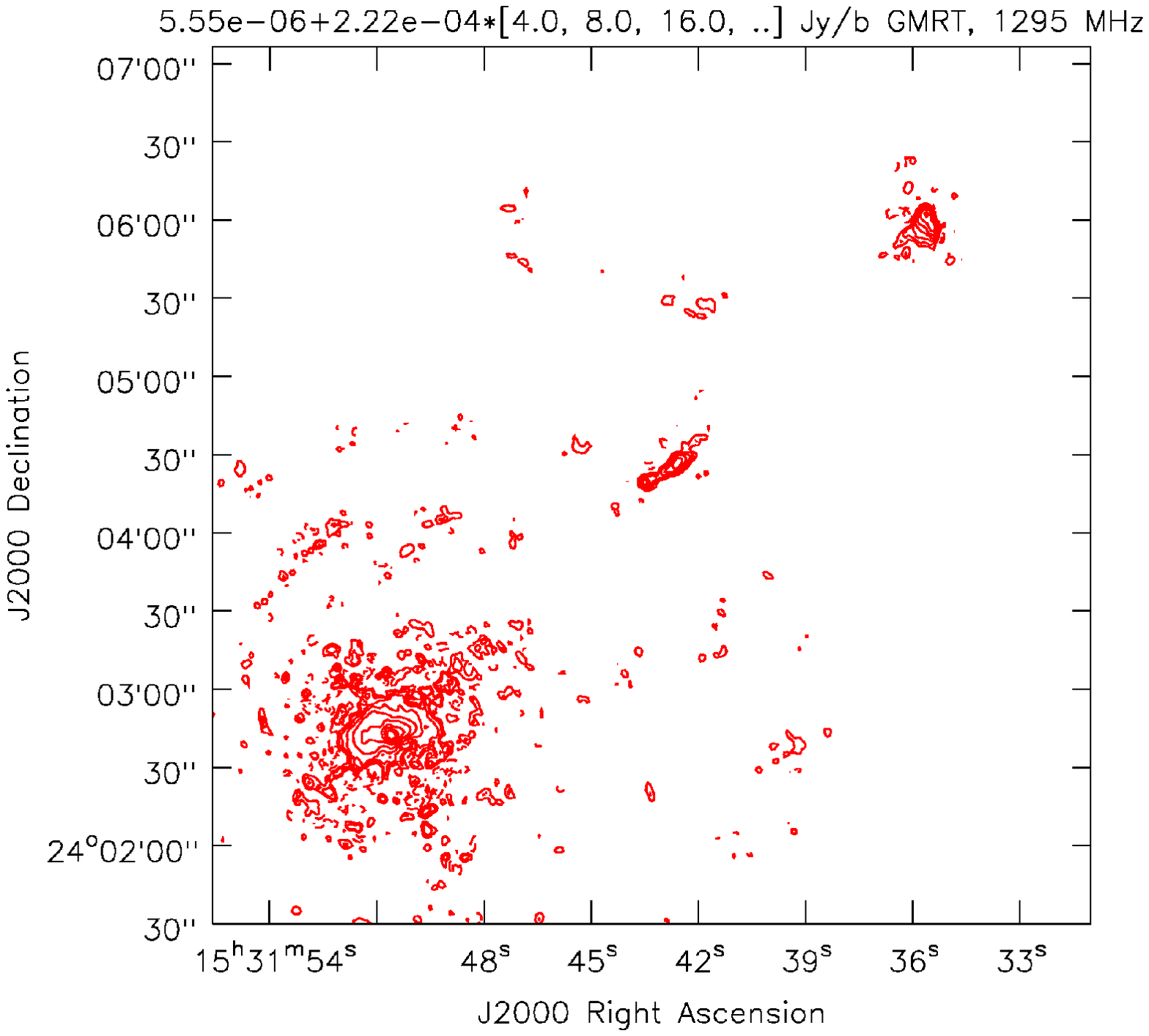}
\hbox{
\includegraphics[width=1.8in,angle=-90]{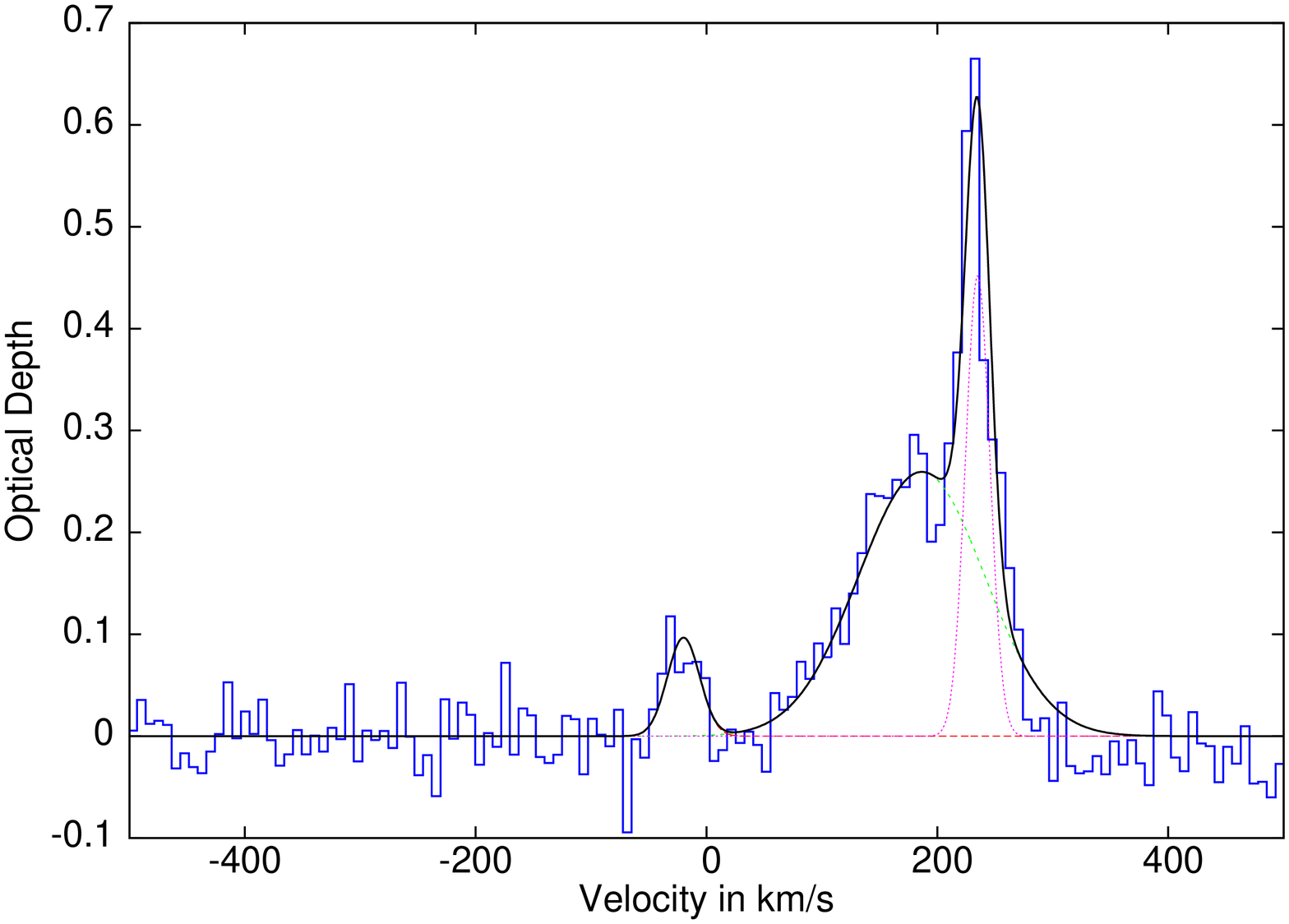}
\includegraphics[width=1.8in,angle=-90]{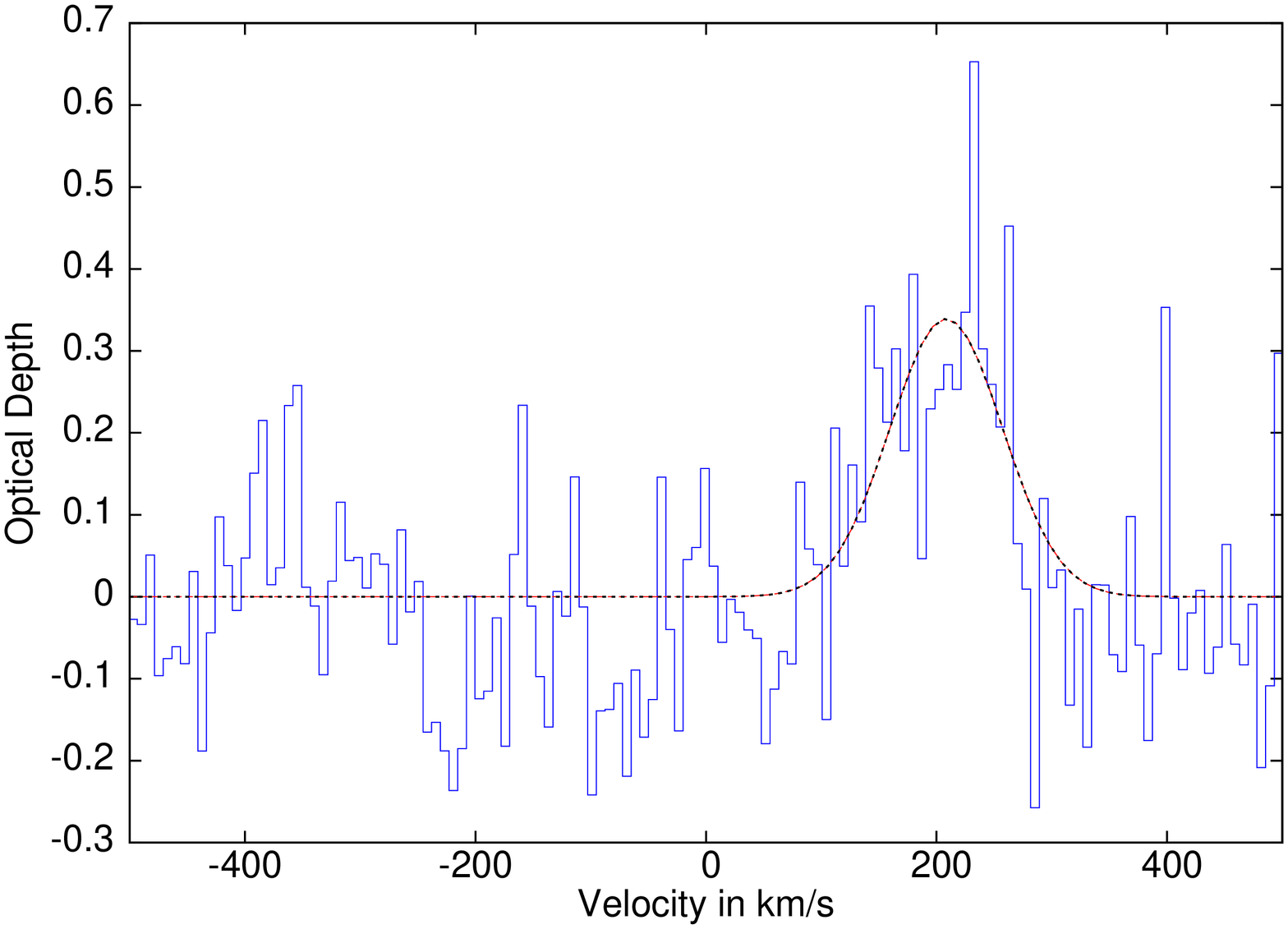}
}
}
\caption[]{Upper panel: GMRT full-resolution image of 3C321 at 1295 MHz with
an angular resolution of 2.50$\times$1.37 arcsec$^2$ along a PA of 71$^\circ$; 
lower panel: Gaussian fits to the 
optical depth vs relative velocity towards the core (left panel) and the knot (right panel) of 
3C321 respectively. The velocity is relative to the optical systemic velocity corresponding to red-shift of 0.0961.}
\label{GF1}
\end{figure}

In Fig.~\ref{GF1} we present the GMRT image of 3C321 at 1295 MHz which was obtained from channels free from absorption. 
This shows the radio core, the knot and a somewhat diffuse jet towards the north-west,
and both the hot-spots on opposite sides of the nucleus.  The peak
flux density of the radio core in our image is  25.2 mJy/beam. The flux densities of
the different features are listed in Table ~\ref{table3}. The emission from the 
north-western (NWL) and south-eastern (SEL) lobes are almost entirely from the hotspots
seen in these images, while the central region (Cen) includes the core and more extended
emission from the jet close to the nucleus which is not well resolved in our lower resolution images. 

The H{\sc i} absorption spectrum towards the core of 3C321 is
presented in the lower panel of Fig.~\ref{GF1}. The H{\sc i} absorption, reported earlier from Arecibo 
observations (Mirabel 1990) with low spatial resolution (3.6 arcmin), is established to be clearly
towards the core of the radio galaxy. There is also evidence of weak absorption towards
the knot in the jet (Fig.~\ref{GF1}).
No absorption has been detected towards the hotspots. The H{\sc i} column density, 
$N$(H{\sc i}), integrated over the entire spectrum towards the core using the relation

\begin{center}
\begin{tabular}{c c l r}

{$N$(H{\sc i})} &=& $1.823\times10^{18}$ {\Large $\frac{{{T}_{\rm s}}~\int{\tau(v)dv}}{f_c}$}~${\rm cm^{-2}}$ & \\ \\
                  &=& $1.93\times10^{18}$ {\Large $\frac{{{T}_{\rm s}}\tau_p\Delta v}{f_c}$}~${\rm cm^{-2}}$ & ~~~~~~~~~~~~~~~~~~~~~~~~~(1) \\
\end{tabular}
\end{center}

\noindent where ${T}_{\rm s}$, $\tau$ and $f_c$ are the spin temperature, optical depth at a velocity
$v$ and the fraction of background emission covered by the absorber respectively, is  
9.23 $\times$10$^{21}$(${T}_{\rm s}$/100)(1.0/$f_c$) cm$^{-2}$. In above equation, $\tau_p$ 
and $\Delta v$ are peak optical depth and FWHM (Full Width at Half Maximum) 
of the gaussians fitted to the absorption profiles respectively. 
We have assumed ${T}_{\rm s}$=100 K and $f_c$=1.0. 
For non-detections, the upper limits on the H{\sc i} column densities were calculated by replacing the $\tau_p$ in 
the above equation by 3$\sigma$ upper limits to the peak optical depths estimated from the rms in the spectra, and
assuming $\Delta v$=100 kms. The absorption profile towards the core has been 
fitted with three Gaussian components (Fig.~\ref{GF1}), and the fit parameters are
summarised in Table~\ref{gauss}. With similar assumptions, the column density towards
the knot in the jet is  7.53 $\times$10$^{21}$(${T}_{\rm s}$/100)(1.0/$f_c$) cm$^{-2}$.
The upper limits to the H{\sc i} column density for the peaks of emission in the NW 
and SE lobes are 8.65 and 1.78 $\times$10$^{20}$(${T}_{\rm s}$/100)(1.0/$f_c$) cm$^{-2}$  respectively. 
All the components of the absorption profile are red-shifted relative to the 
optical systemic velocity, except for the  weak feature at $-$19 km s$^{-1}$. 
Although this feature is  relatively weak, there is 
some evidence of it in the Arecibo spectrum (Mirabel 1990).

Kinematics of the gas in the circum-nuclear regions of the host galaxies
may sometimes lead to different velocities for different emission lines, 
making it difficult to infer precisely the kinematics of the absorbing cloud 
(e.g. Tadhunter et al. 2001; Vermeulen et al. 2006). In the case of 3C321,
the H{\sc i} absorbing clouds appear significantly red-shifted by 
over $\sim$200 km s$^{-1}$. Considering the cores of extended radio sources with
H{\sc i} absorption, peaks of absorption are known to be red-shifted, blue-shifted
or consistent with the velocity of the emission line gas (e.g. Conway \& Blanco
1995; Morganti et al. 2001; Gupta \& Saikia 2006a; Chandola et al., in preparation), 
similar to what has been
seen in the case of CSS and GPS objects (Vermeulen et al. 2003; Gupta et al. 2006).
In addition, fast and broad outflows in H{\sc i} with velocities extending up to
several thousand km s$^{-1}$ have been reported for a number of radio galaxies
(Morganti, Tadhunter \& Oosterloo 2005; Emonts et al. 2005), possibly due to
outflowing gas caused by jet-cloud interactions. The blue-shifted tail seen in
3C452 (Gupta \& Saikia 2006a), is perhaps caused by interaction of the jet with the absorbing gas. The
redshift-components towards the core in 3C321 suggest that these might be due 
to infalling gas.   
%%%%%%%%%%%%%%%%%%%%%%%%%%%%%%%%%%%%%%%%%%%%%%%%%%%%%%%%%%%%%%%%%%%%%%%%%%%%%%%%%%%%%%%%%%%%%%%%%%%%%%%%

\begin{figure}
\vbox{
\hbox{
\includegraphics[width=1.8in,angle=-90]{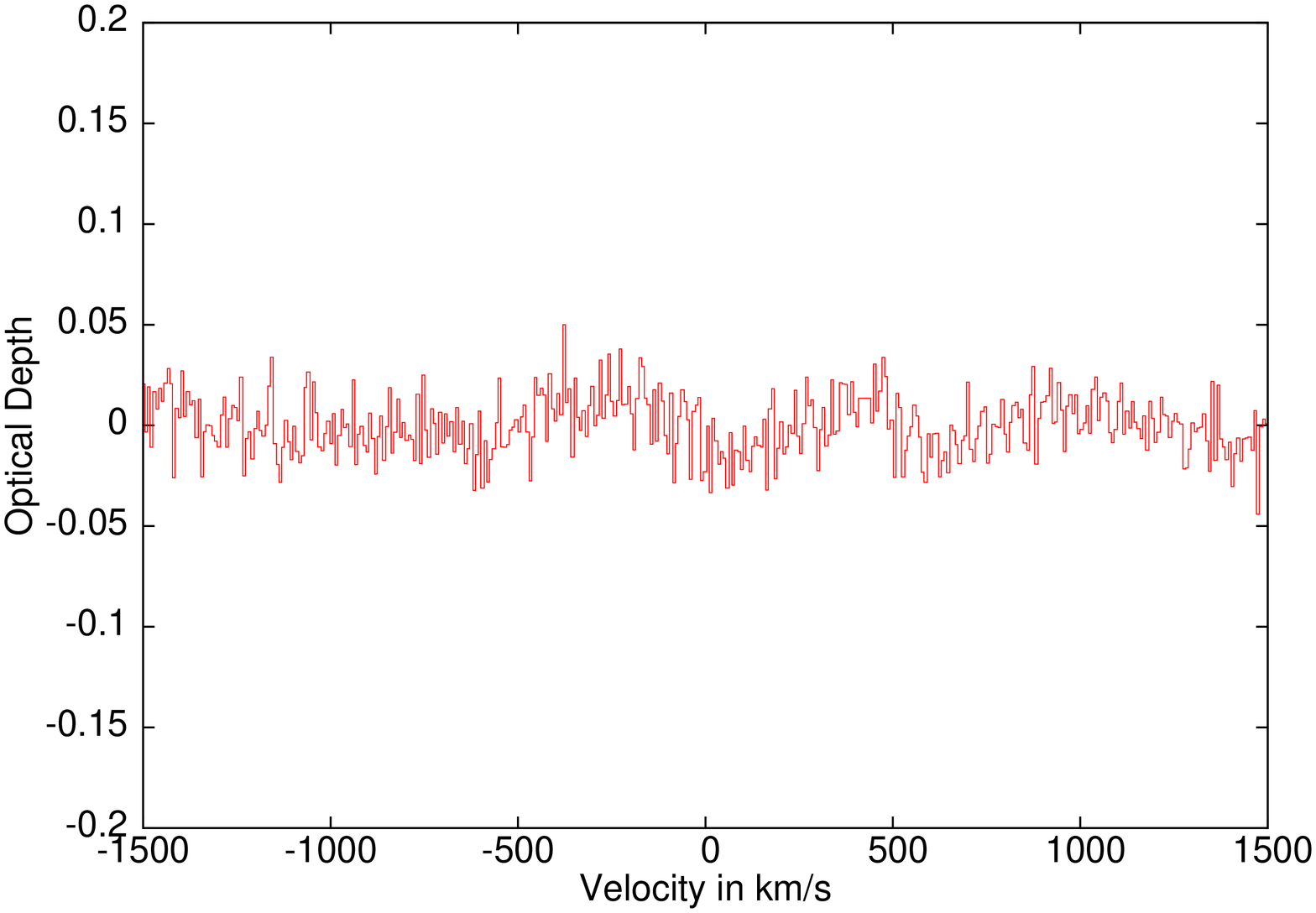}
\includegraphics[width=1.8in,angle=-90]{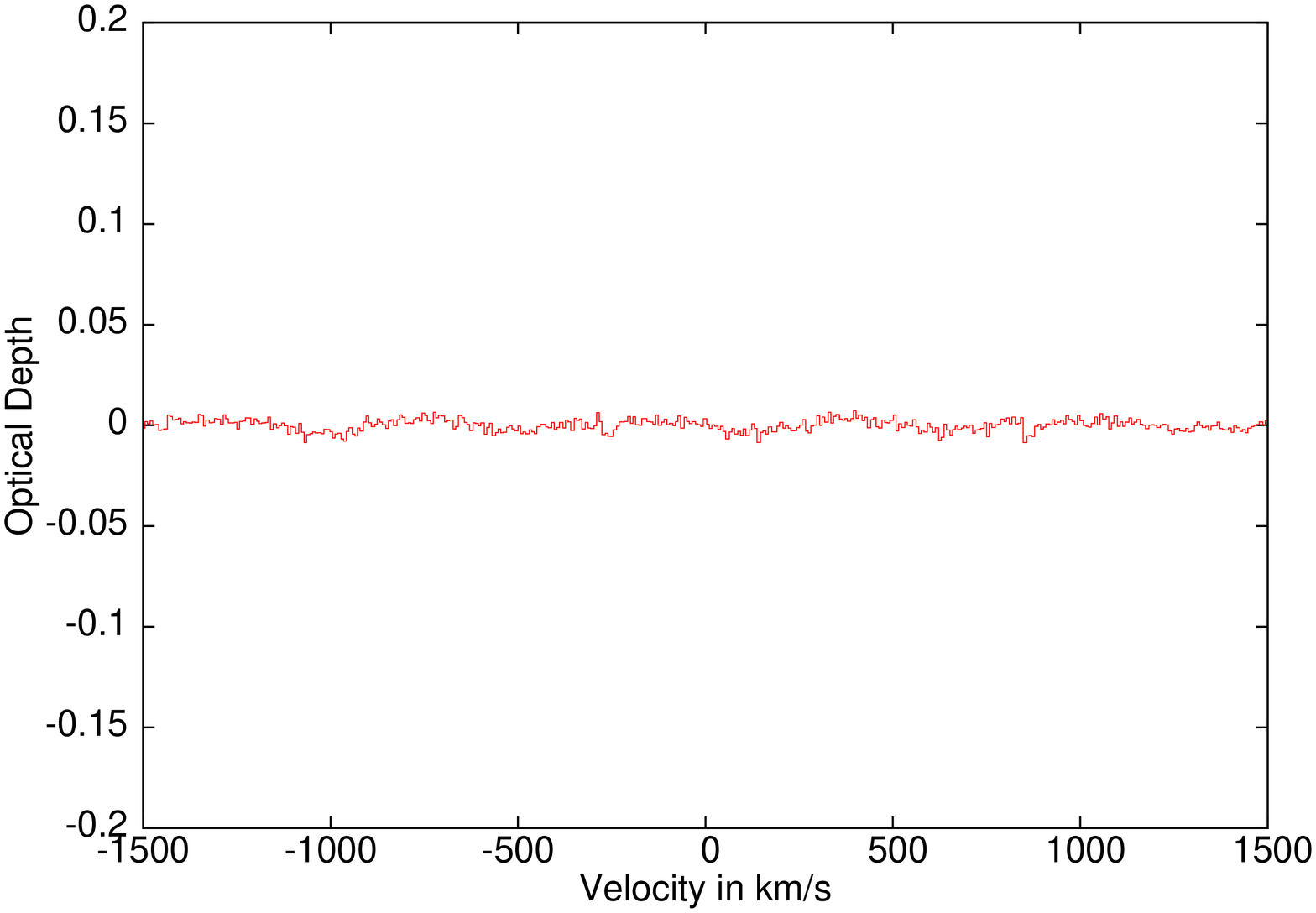}
}
}
\caption[]{Spectra plotted as optical depth vs relative velocity towards the NW (north western) 
and SE (south eastern) lobes of 3C321. The velocity is relative to the optical systemic velocity corresponding to a 
red-shift of 0.0961.} 
\label{GF2}
\end{figure}

%%%%%%%%%%%%%%%%%%%%%%%%%%%%%%%%%%%%%%%%%%%%%%%%%%%%%%%%%%%%%%%%%%%%%%%%%%%%%%%%%%%%%%%%%%%%%%%%%%%%%%%%%

\begin{table}
\caption{Multiple Gaussian fit to the H{\sc i} absorption spectrum towards the core and the knot of 3C321.}
\begin{center}
\begin{tabular}{ c l l c c }
\hline
Id. &   Velocity  &  FWHM & $\tau_{p}$   & $N$(H{\sc i}) \\
no. &    $v$      & $\Delta v$      &  & 10$^{20}$($\frac{{T}_{\rm s}}{100})(\frac{f_c}{1.0})^{-1}$\\
    &   km s$^{-1}$ &km s$^{-1}$ &       & cm$^{-2}$ \\
\hline
\multicolumn{5}{c}{\bf core}                           \\
1$^\ast$ &  $-$19.9(3.8) & 32.9(9.0)& 0.097(0.023) & 6.1(3.1) \\
2   &  186.4(3.9) & 128.8(7.1)& 0.259(0.013) & 64.5(6.8)\\
3   &  235.0(0.7) & 24.9(2.0) & 0.452(0.029) & 21.7(3.2) \\
\hline
\multicolumn{5}{c}{\bf knot}                           \\
1   &  208.5(7.5)& 115.2(17.6)& 0.339(0.045) & 75.4(21.5) \\
\hline
\end{tabular}
\end{center}

$^\ast$This feature is weak. However, there appears to be a similar weak feature in the Arecibo spectrum (Mirabel 1990). 
\label{gauss}
\end{table}
%%%%%%%%%%%%%%%%%%%%%%%%%%%%%%%%%%%%%%%%%%%%%%%%%%%%%%%%%%%%%%%%%%%%%%%%%%%%%%%%%%%%%%%%%%%%%%%%%%%%%%%%% 
%%%%%%%%%%%%%%%%%%%%%%%%%%%%%%%%%%%%%%%%%%%%%%%%%%%%%%%%%%%%%%%%%%%%%%%%%%%%%%%%%%%%%%%%%%%%%%%%%%%%%%%%%%%%%%%%%%%%%%

\begin{figure}
\vbox{
\includegraphics[width=3.0in,angle=0,viewport=15 212 573 627, clip]{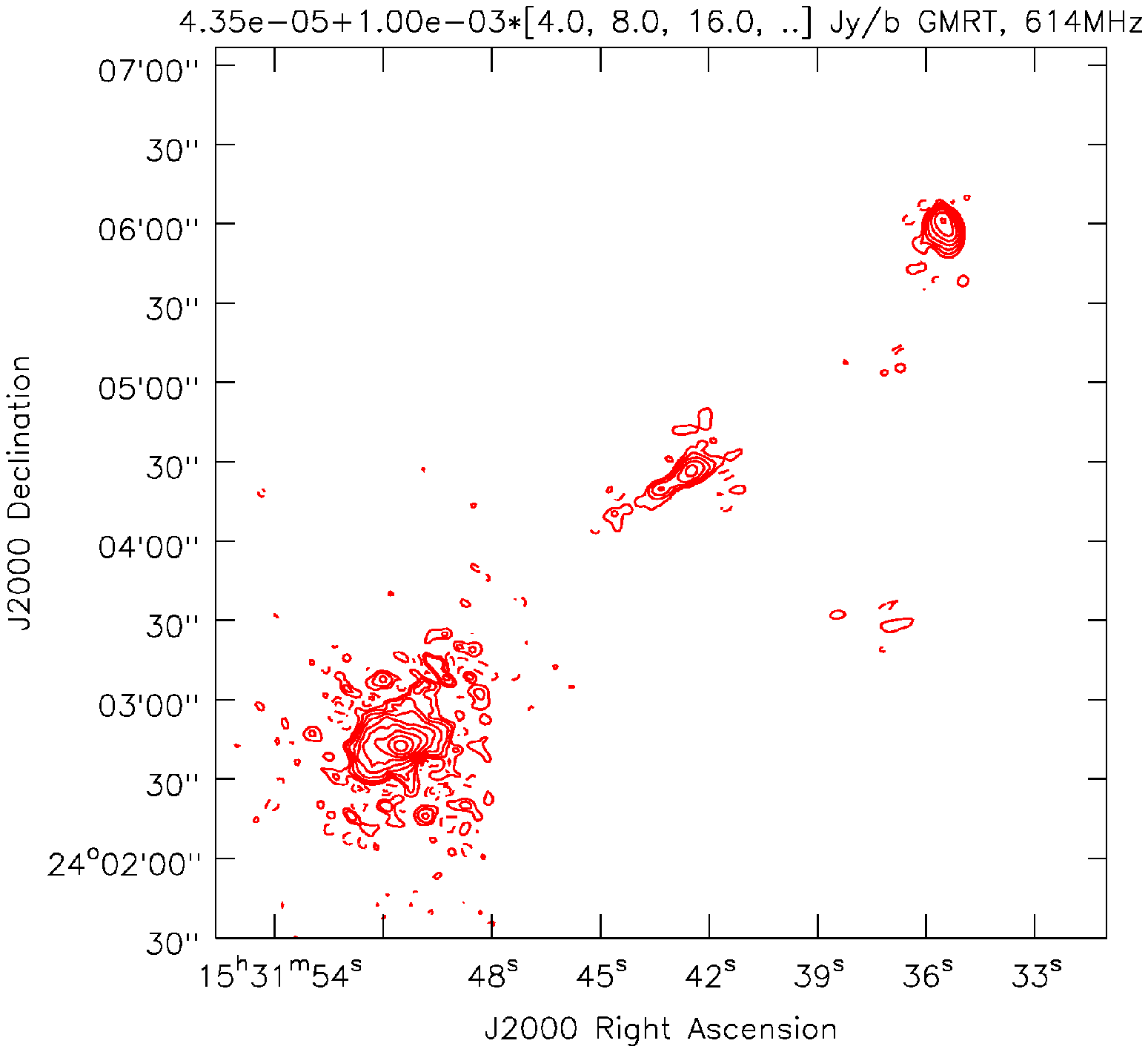}
\hskip -0.8cm
\includegraphics[width=3.0in,angle=0,viewport=19 212 573 627, clip]{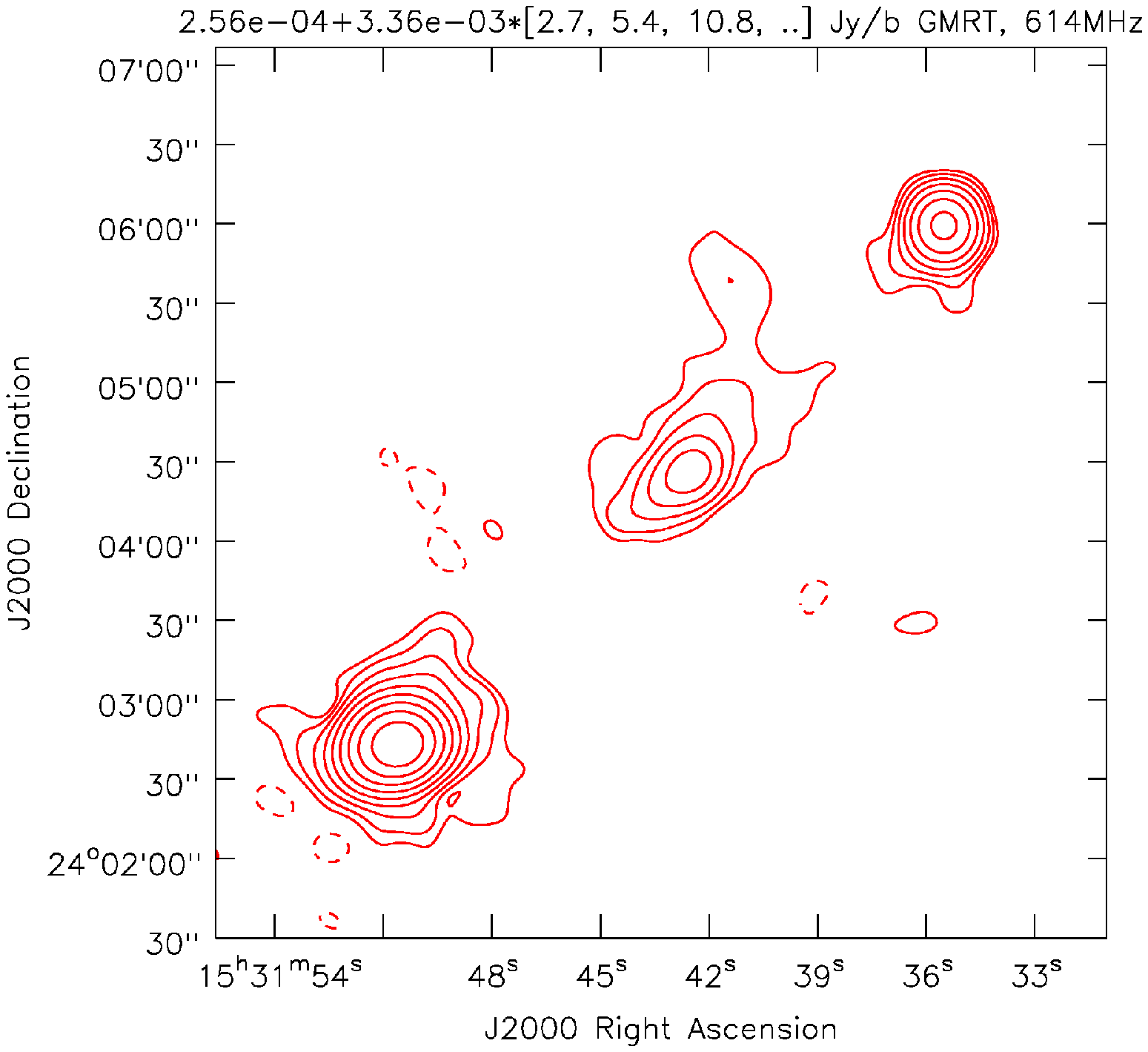}
\includegraphics[width=3.0in,angle=0,viewport=19 212 573 627, clip]{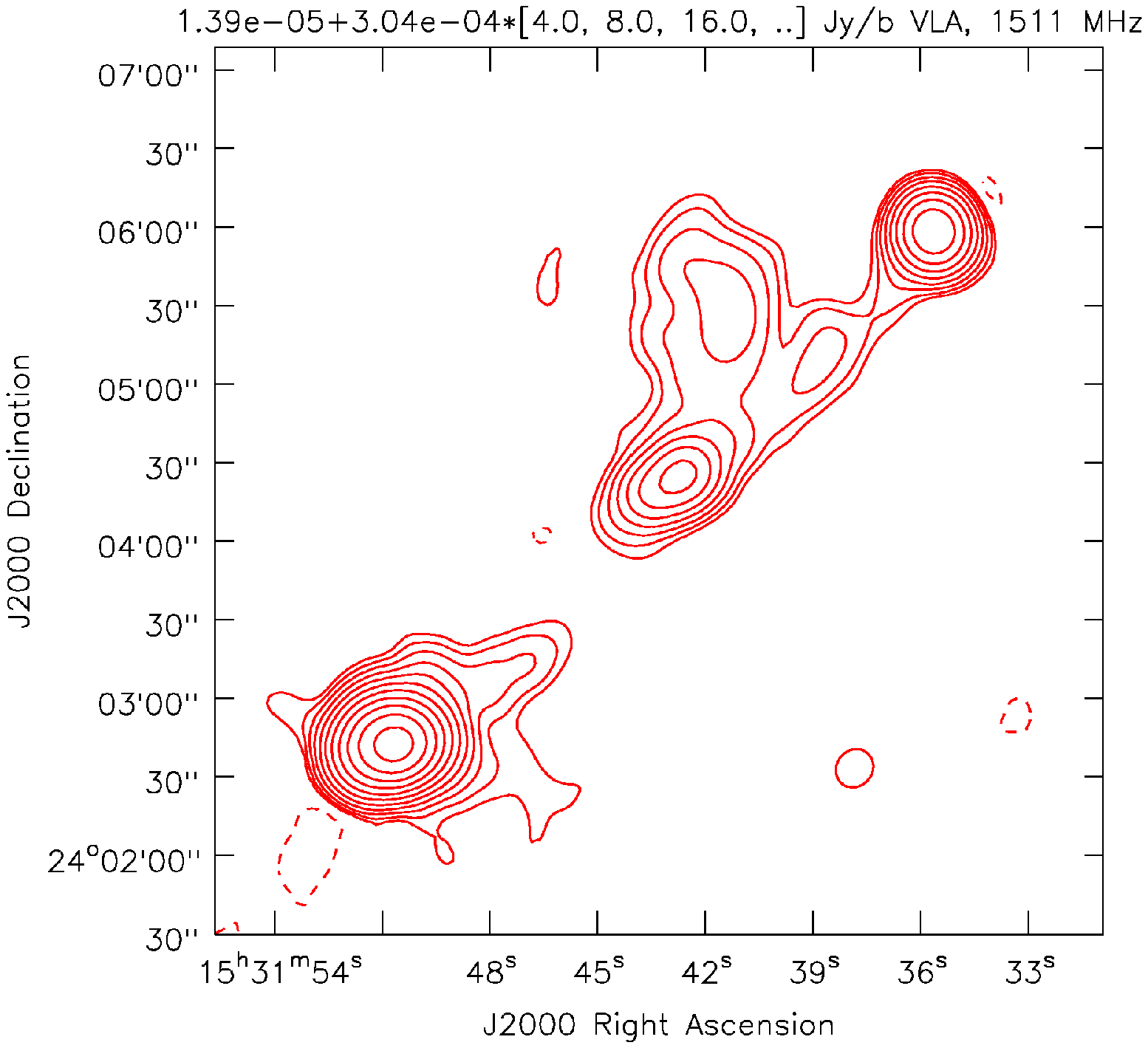}
\hskip -0.8cm 
\includegraphics[width=3.0in,angle=0,viewport=19 212 573 627, clip]{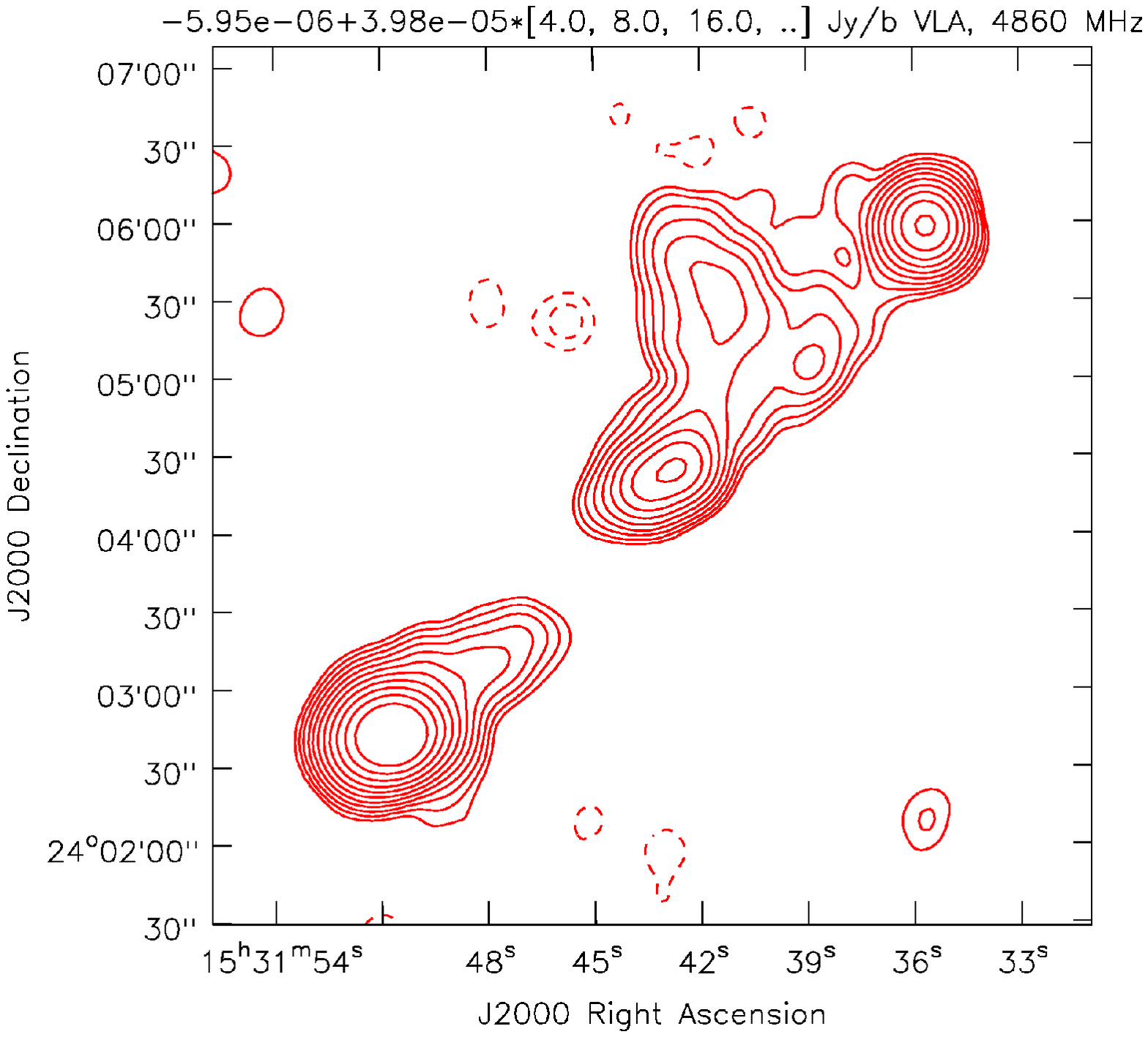}
}
\caption[]{Upper panels: GMRT images at 614 MHz with an angular resolution of 4.48$\times$3.69 arcsec$^2$ 
along PA=59.1$^\circ$ arcsec (left) 
and convolved to a resolution of 16 arcsec (right).  Lower panels: VLA images at 1511 and 4860 MHz
with an angular resolution of 16 arcsec. } 
\label{CMAPS}
\end{figure}

%%%%%%%%%%%%%%%%%%%%%%%%%%%%%%%%%%%%%%%%%%%%%%%%%%%%%%%%%%%%%%%%%%%%%%%%%%%%%%%%%%%%%%%%%%%%%%%%%%%%%%%%

\subsection{Radio images}
\begin{figure}
\vbox{
\hbox{
\includegraphics[width=2.6in,angle=0]{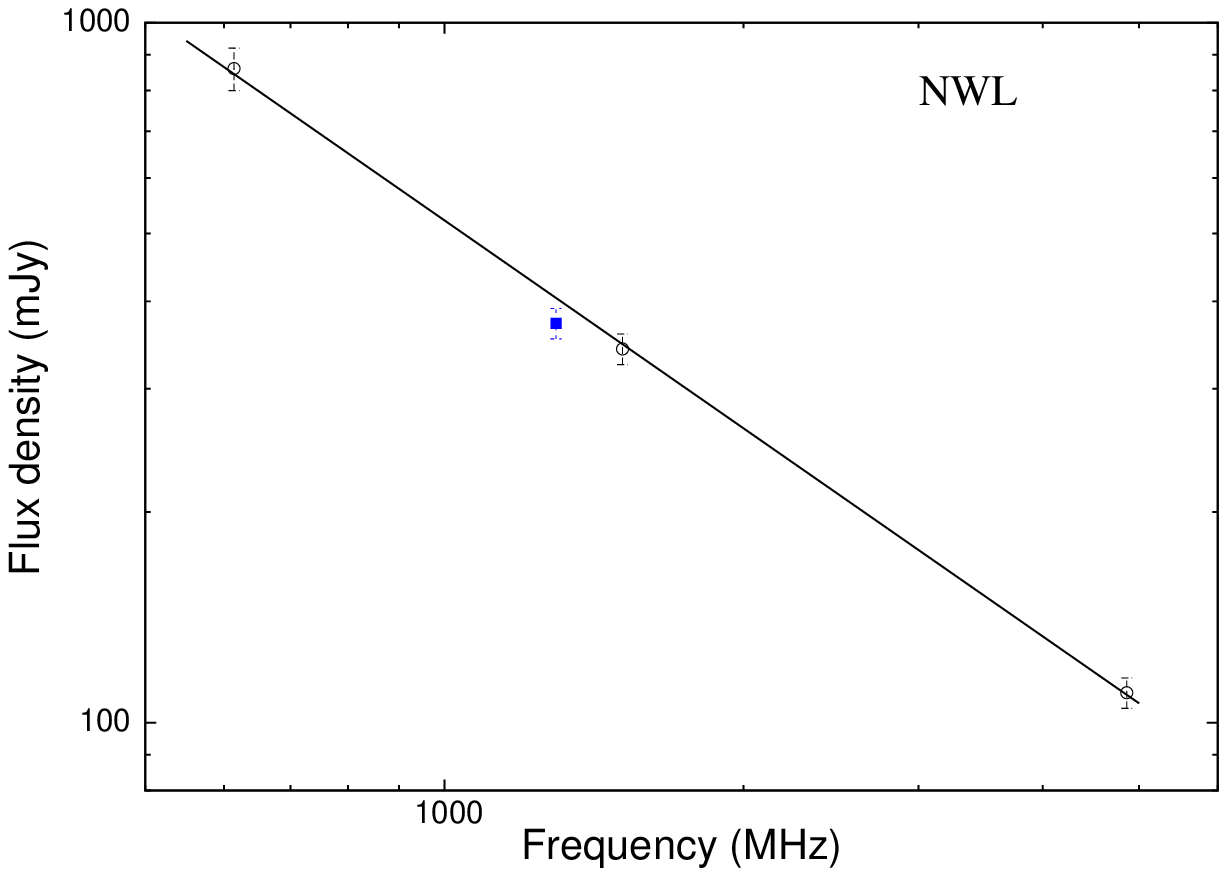}
\includegraphics[width=2.6in,angle=0]{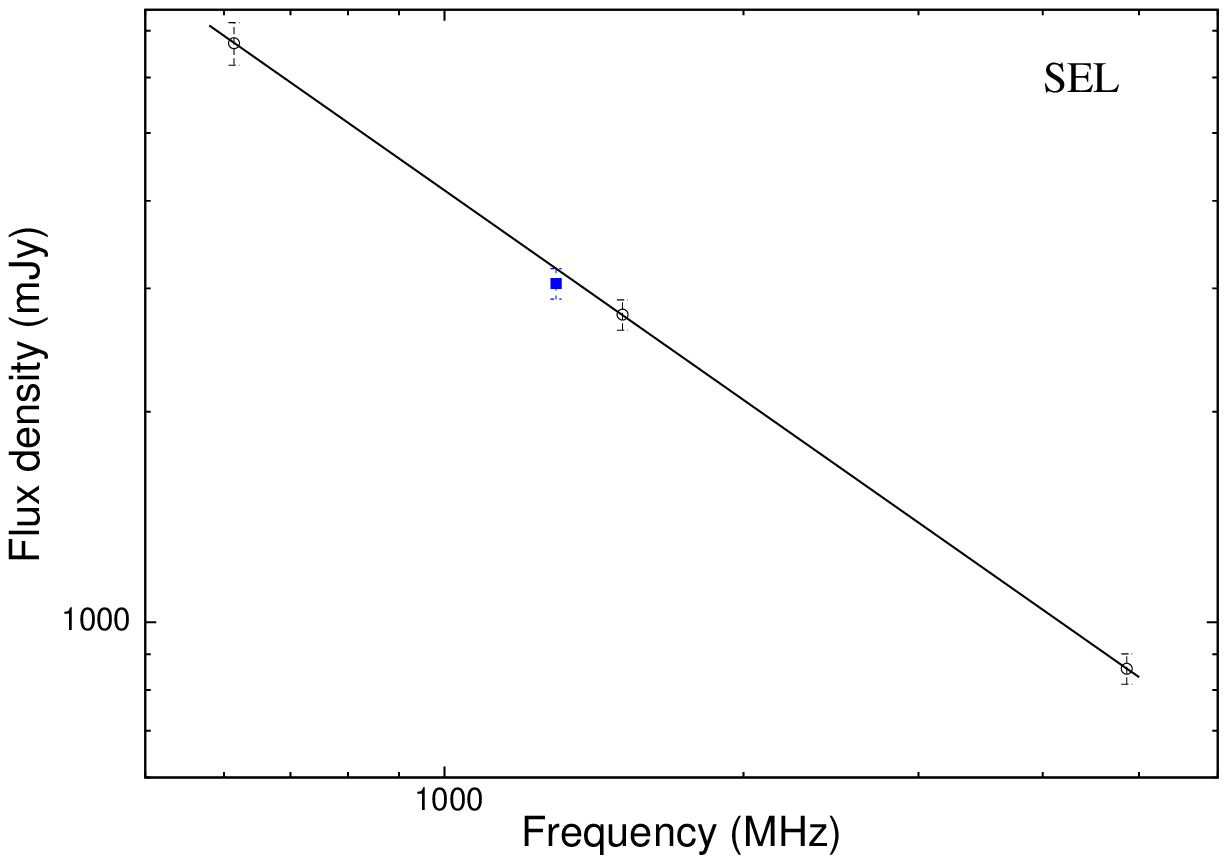}
}
\hbox{
\includegraphics[width=2.6in,angle=0]{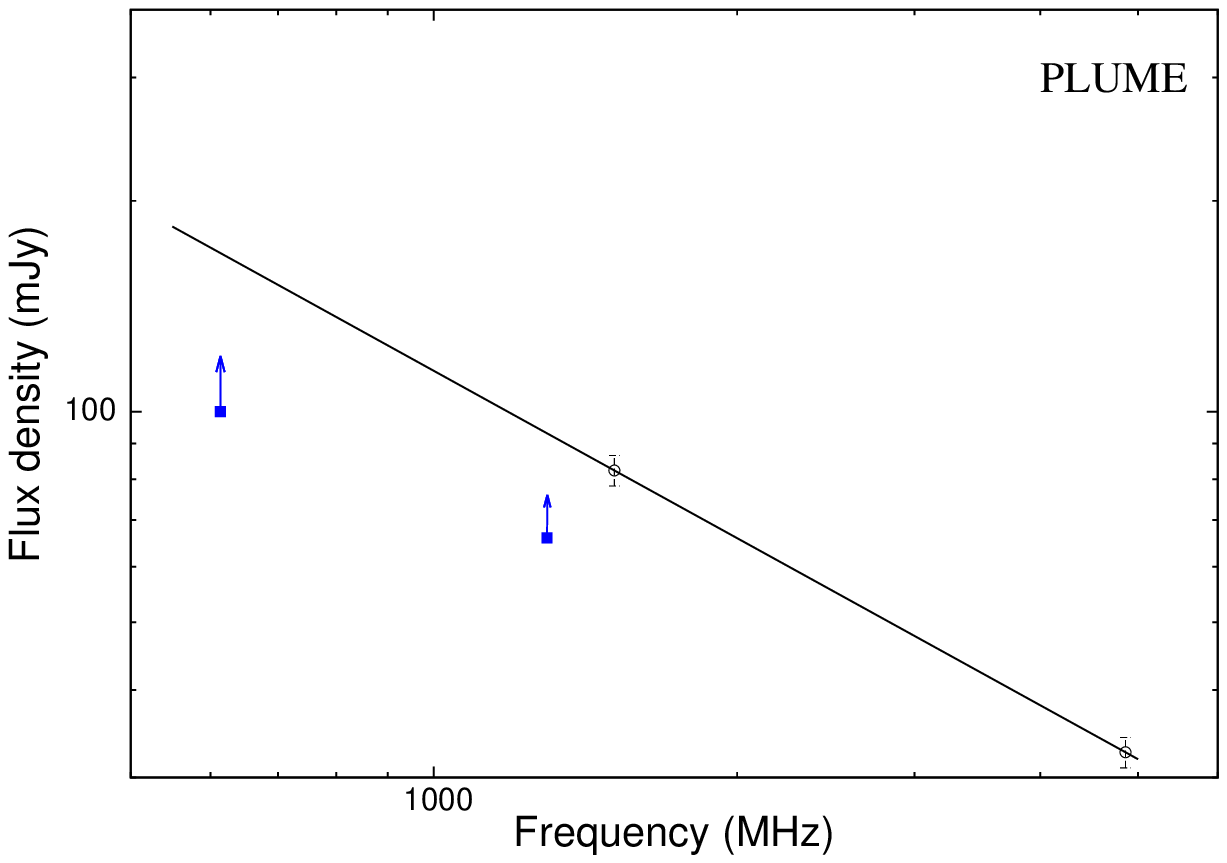}
\includegraphics[width=2.6in,angle=0]{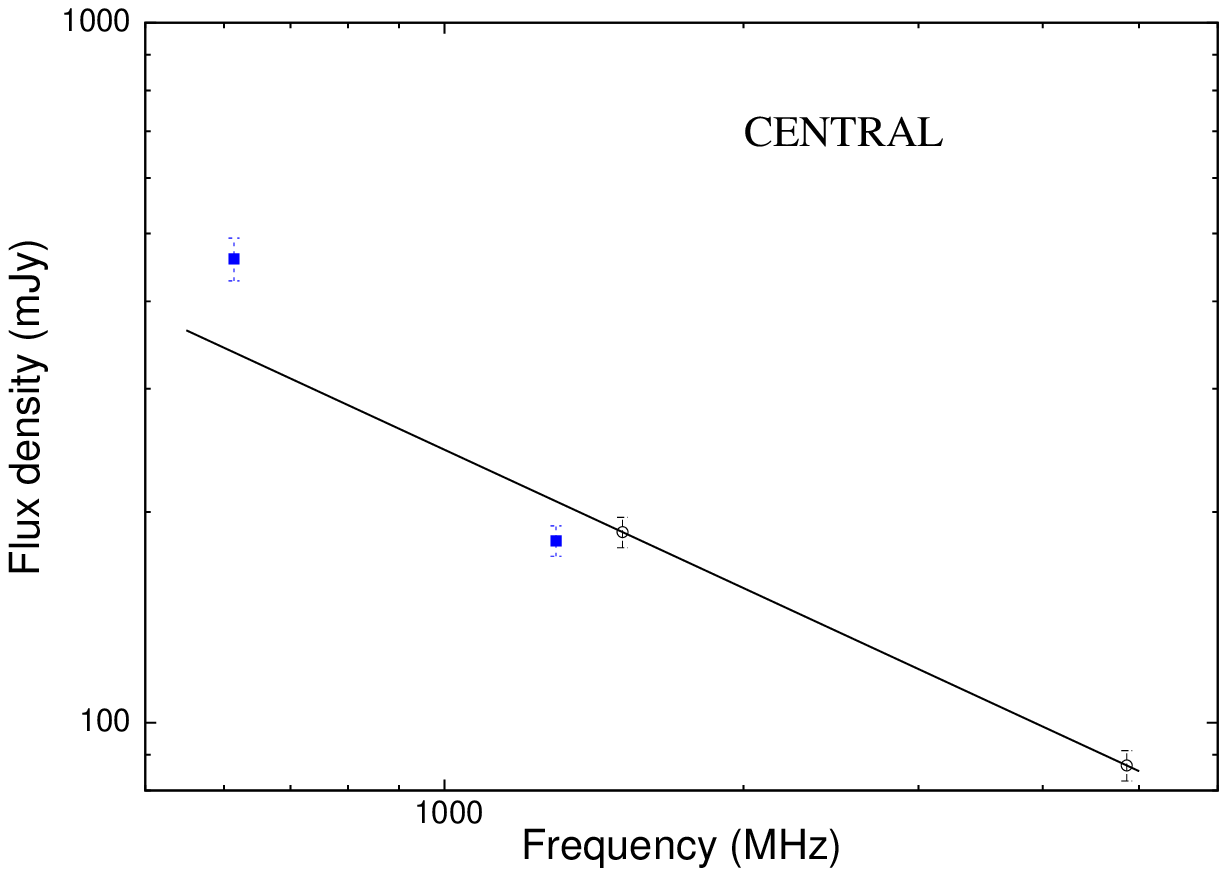}
}
}
\caption[]{Figure shows the spectra of the various components namely the 
NWL (north western lobe), the SEL (south eastern lobe), the plume
and the central region which includes the core. (Points shown as blue squares have not 
been used while fitting for spectral index. Arrows show lower limits to
the flux density estimates.)} 
\label{SD1}
\end{figure}

The GMRT images at 614 MHz and the VLA images at 1511 and 4860 MHz are shown in Fig.~\ref{CMAPS}. 
The flux densities estimated from the images with an angular resolution of 16 arcsec
are presented in Table~\ref{table3}.
The errors in the flux densities are approximately 
7 per cent at 614 MHz and 5 per cent at higher frequencies. These errors are mainly flux density scale uncertainities
and calibration errors. All other errors are negligible. Spectral index error due to error propagated from these
uncertainities is 0.13. The spectra of the different components are shown in Fig.~\ref{SD1}.  
The lobes and the central region are seen clearly in all the images. The
plume of emission is best seen in the VLA images at 1511 and 4860 MHz. The 
overall extent is smaller in the GMRT 614-MHz image.
The spectral index ($S(\nu) \propto \nu^{-\alpha}$) of both lobes between $\sim$600 and 5000 MHz
is $\sim$ 0.99.
The spectra of the lobes appear straight, and assuming that
the break frequency is above 5 GHz, we can estimate an upper limit to their spectral ages. 
Assuming a proton to electron energy ratio of unity, and integrating from 10 MHz to 100 GHz, 
using the standard equations (e.g. Miley 1980) yields an equipartition magnetic field of 
$\sim$8.0 $\mu$G for both the lobes, suggesting a spectral age of $\lapp$26 Myr for the lobes. 
Detection of X-ray emission from the hot-spots, which is likely to be of synchrotron
origin, suggests that these are young features, continuously being supplied with
relativisitic plasma from the nucleus (Hardcastle et al. 2004; Evans et al. 2008).

The spectral index of the central region between $\sim$1500 and 5000 MHz is 0.66. 
The flux density of the central region appears to flatten towards higher frequencies.
This is due to the increased contribution of the radio core which is not well resolved
in our low resolution observations. The core has a flux density of 30 mJy at 5 GHz (Giovannini et al. 1988)
while our GMRT observations at 1295 MHz yield a peak flux density of 25.2 mJy beam$^{-1}$.
Subtracting these values from the total flux densities of the central regions gives a 
straight spectrum between $\sim$600 and 5000 MHz with a spectral index of 0.98. 
%%%%%%%%%%%%%%%%%%%%%%%%%%%%%%%%%%%%%%%%%%%%%%%%%%%%%%%%%%%%%%%%%%%%%%%%%%%%%%%%%%%%%%%%%%%%%%%%%%%%%%%%%%%%%%%%%%%%%%%%%
%%%%%%%%%%%%%%%%%%%%%%%%%%%%%%%%%%%%%%%%%%%%%%%%%%%%%%%%%%%%%%%%%%%%%%%%%%%%%%%%%%%%%%%%%%%%%%%%%%%%%%%%%%%%%%%%%%%%%%%%%
\begin{table}
\caption{The flux densities of the components. 
Column 1: The frequency used in making the images;
columns 2 to 6: flux density of 
the different components and the total flux density of the source estimated
from images with an angular resolution of 16 arcsec except for 1295 MHz which
has angular resolution of 2.50$\times$1.37 arcsec$^2$ along a PA of 71$^\circ$.}
\label{table3}
\begin{center}
\begin{tabular}{c c c c c c  c }
\hline
 Obs.    &  NWL   & SEL     &  Plume & Cen   & Total     \\
 Freq.   &        &         &        &           &           \\
(MHz)    & (mJy)  & (mJy)   &  (mJy) &  (mJy)    & (mJy)     \\
 (1)     &  (2)   &  (3)    &  (4)   &  (5)     &  (6)      \\
\hline
 614     &  860   &   6718  &$\sim$100$^\ast$&  460   &   8197     \\
 1295    &  372   &   3048  &$\sim$66$^\ast$ &  182   &  3686     \\
 1511    &  342   &   2751  &   82        &  187      &  3384     \\
 4860    &  110   &    858  &   33        &   87      &  1096     \\
\hline
\end{tabular}
\end{center}

{$^\ast$ Flux density has been estimated over same area as the VLA images, and the total flux 
density of the plume is likely to be higher}.   
\end{table}
%%%%%%%%%%%%%%%%%%%%%%%%%%%%%%%%%%%%%%%%%%%%%%%%%%%%%%%%%%%%%%%%%%%%%%%%%%%%%%%%%%%%%%%%%%%%%%%%%%%%%%%%%%%%%%%%%%%%%%%%%

\subsubsection{The plume of emission}
The spectrum of the plume between 1511 and 4860 MHz, estimated from similar
resolution VLA images, is steep  with a spectral index of $\sim$0.8. The flux 
density of the plume at 614 MHz has been estimated over same area as at
the higher frequencies. The plume is seen over a smaller extent at 614 MHz
compared with the VLA images, and our estimate of the flux density at this
frequency is $\sim$100 mJy. 
Lal, Hardcastle \& Kraft (2008) estimate the flux density of the plume which
they term the `north wing' to be 32.1 and 17.2 mJy from their
high- and low-resolution images at 610 MHz, and a spectral index of 0.30
between 1500 and 610 MHz. Our flux density estimate is clearly significantly higher than
those of Lal et al. We imaged the archival GMRT data at the lower frequencies to determine
the spectrum of the plume, but the images were not of satisfactory quality to 
give reliable estimates of the flux densities of the plume (see for example the
240-MHz image of Lal et al. 2008). The plume is also barely seen in the TIFR GMRT 
Sky Survey (TGSS; Sirothia et al., in preparation) image at 148 MHz; clearly better quality 
data are required to determine the low-frequency spectrum of the plume reliably.  
This would help determine any break in the spectrum and explore the possibility of whether 
the plume might be relic emission due to an earlier cycle of activity. Although the
present suggestion is that the plume is caused by the deflection of the jet 
(e.g. Evans et al. 2008), the detailed spectrum of the plume would help examine
the alternate possibility of this being relic emission. If the plume of emission is 
indeed due to an earlier cycle, the detection of H{\sc i} in absorption would be consistent 
with the trend for a high incidence of absorption in sources with evidence of recurrent activity (cf. Saikia \& Jamrozy
2009 for a review; Salter et al. 2010 and references therein).

\section{Concluding remarks}
The GMRT H{\sc i} observations show that the absorption occurs towards the core component,
with also a suggestion of absorption towards the knot in the jet 6 kpc north-west 
of the nucleus. This would indicate the size of the absorber to be $\gapp$ 6 kpc
if it is a single absorber. The absorption profile towards the core has multiple 
components with a total column density of 
$N$(H{\sc i})=9.23$\times$10$^{21}$(${T}_{\rm s}$/100)($f_c$/1.0) cm$^{-2}$. The column
density towards the knot is 
$N$(H{\sc i})=7.54$\times$10$^{21}$(${T}_{\rm s}$/100)($f_c$/1.0) cm$^{-2}$. No absorption is
seen towards the hot-spots.

Radio continuum observations of lobes with similar resolution show  that their
spectra are straight up to 5 GHz, suggesting spectral ages of $\lapp$26 Myr 
for the lobes. The detection of X-rays from the hot-spots suggests that these are being continuously
fed by the jets, consistent with their spectral ages. The full extent of the 
plume is not seen in our 614-MHz GMRT and also barely seen in the TGSS image at
148 MHz. However, better quality data are required to determine the spectrum of
the plume reliably over a large frequency range to try and examine a possible break
in the spectrum and explore whether the plume might be relic emission from an
earlier cycle of activity.

\section*{Acknowledgments} 
We thank George Privon for useful discussions and sharing his results on 3C321
with us, Dave Green, the reviewer for many useful comments which helped improve
the presentation of the paper, and the staff of GMRT for their help with the
observations. The GMRT is a national facility operated by the
National Centre for Radio Astrophysics of the Tata Institute of Fundamental Research. 
We thank the NRAO for use of the archival data.
The National Radio Astronomy Observatory  is a facility of the National Science Foundation
operated under co-operative agreement by Associated Universities Inc. 
This research has made use of the NASA/IPAC extragalactic database (NED) which is operated
by the Jet Propulsion Laboratory, Caltech, under contract with the National Aeronautics
and Space Administration. 

{}
\end{document}